%% file: wpcf_jmargutti.tex
\documentclass[WPCF,manyauthors]{wpcfTemplate}

\usepackage{fixltx2e}
\usepackage{float}
\usepackage{rotating}
\usepackage{graphicx}
\usepackage{subcaption}
\begin{document}%
%
%
\begin{titlepage}
%
\title{The search for magnetic-induced charged currents in Pb--Pb collisions with ALICE}
\ShortTitle{Search for magnetic-induced charged currents with ALICE}   
\author{Jacopo Margutti$^{1,2}$ (for the ALICE Collaboration)}
\author{
1. Utrecht University, Domplein 29, 3512 JE Utrecht, Netherlands \\
2. Nikhef, Science Park 105, 1098 XG Amsterdam, Netherlands
}
\author{Email: jacopo.margutti@cern.ch}
\ShortAuthor{XII$^{th}$ edition of WPCF}      
\begin{abstract}
In non-central heavy-ion collisions unprecedented strong magnetic fields, of the order of 10$^{18}$ Gauss, are expected to be produced. The interplay of such fields with QCD anomalies in the Quark--Gluon Plasma (QGP) has been predicted to lead to a number of interesting phenomena, such as the Chiral Magnetic Effect (CME). While several experimental observations are partially consistent with predictions of a CME signal, it is hard to distinguish them unambiguously from a combination of more mundane phenomena present in the anisotropic expansion of the QGP. This makes it imperative to establish that the early-time magnetic field has observable consequences not related to the anomalous QCD effects on final-state charged particles and to calibrate its strength. We test a recent prediction of a pure electromagnetic effect, which may arise in heavy-ion collisions. The varying magnetic field would induce a current within the QGP, which is expected to leave a very peculiar imprint on final-state particles: a contribution to directed flow which is asymmetric both in charge and pseudorapidity. We report the measurement of the charge dependence of directed flow with respect to the spectator plane for unidentified charged particles in Pb--Pb collisions at $\sqrt{s_\text{NN}} = 5.02$ TeV.
\end{abstract}
\end{titlepage}
\input{main.tex}               

\end{document}

%% file: main.tex
\section{Introduction}

In non-central heavy-ion collisions unprecedented strong magnetic fields are expected to be produced. The movement of the spectator (non-interacting) protons, according to the Biot-Savart law, would induce a strong (up to 10$^{18}$ Gauss at LHC energies) although short-lived ($\sim 1$ fm/$c$) magnetic field in the nuclei overlap region \cite{Tuchin:2013ie}. The interplay of such field with quantum anomalies possibly present in the Quark-Gluon Plasma (QGP) has been predicted to lead to a number of novel phenomena, such as the Chiral Magnetic Effect (CME) \cite{Kharzeev:2007jp}. While several experimental observations are partially consistent with such an effect, it is hard to distinguish them unambiguously from a combination of more mundane phenomena present in the anisotropic expansion of the QGP. This makes it imperative to establish that the early-time magnetic field has observable consequences on final-state charged particles and to calibrate its strength and lifetime.

One possible way to address the issue, which allows disentangling the measurement of the magnetic field from exotic effects such as the CME, is to look at the electromagnetic currents that such field would induce in the QGP. The idea has first been proposed in \cite{Gursoy:2014aka}, together with an order-of-magnitude estimate of the phenomenon. The observable consequence of such currents would be a contribution to the pseudorapidity-odd component of directed flow with opposite sign for opposite charges. Two competing effects were predicted to play a role: the electric fields generated according to Faraday’s law by the decreasing magnetic field and the Lorentz force that charge carriers experience.

\section{Analysis}
The sample of Pb--Pb collisions used for this measurement was recorded with the ALICE detector \cite{ALICE} in November 2015, during the Run 2 of the LHC, at a center of mass energy per nucleon of $\sqrt{s_\text{NN}} = 5.02$ TeV. About $2 \times 10^7$ events (corresponding to an integrated luminosity of 8 $\mu$b$^{-1}$) passed the selection criteria, which include coincidence of signals between two forward detectors (V0A and V0C), a reconstructed primary vertex position along the beam direction within $\pm 10$ cm of the nominal interaction point and the collision centrality within 5--40\%. 
Charged particle tracks with transverse momentum $p_\text{T} > 0.2$ GeV/$c$ and pseudorapidity $|\eta|< 0.8$ are used in this analysis. These tracks are reconstructed using combined information from the Inner Tracking System (ITS) and the Time-Projection Chamber (TPC). The spectator plane is reconstructed from the transverse asymmetry of energy deposited by spectator neutrons in 2x4 segments of two neutron Zero-Degree Calorimeters (ZDCs) located at $\eta > 8.8$ (ZDC-A) and $\eta < -8.8$ (ZDC-C). For each ZDC a flow vector is constructed as
$$ \vec{Q}_\text{A,C} = \frac{ \sum_{j=1}^4 \vec{n}_j \, {E_\text{A,C j}}^\alpha }{ \sum_{j=1}^4 {E_\text{A,C j}}^\alpha } ,$$
where $E_j$ is the energy measured by the $j$-th segment and $\vec{n}_j$ the position of its center (in cm)
$$ n_{x}[4] = \lbrace -1.75, 1.75, -1.75, 1.75 \rbrace, \quad n_{y}[4] = \lbrace -1.75, -1.75, 1.75, 1.75 \rbrace.$$
The parameter $\alpha$ compensates for saturation effects in case of high energy deposition in the calorimeters and in this analysis is set to $\alpha = 0.395$, according to previous Monte Carlo studies \cite{Gemme:2006}.
The directed flow is then measured using the scalar product method \cite{Voloshin:2008dg}
$$ v_1\lbrace \text{ZDC-A,ZDC-C} \rbrace = \frac{ \langle \vec{q} \cdot \vec{Q}_\text{A,C} \rangle }{\sqrt{ | \langle \vec{Q}_\text{A} \cdot \vec{Q}_\text{C} \rangle | } } = \frac{ \langle q_x Q_{\text{A,C}x} + q_y Q_{\text{A,C}y}  \rangle }{\sqrt{ | \langle Q_{\text{A}x} Q_{\text{C}x} + Q_{\text{A}y} Q_{\text{C}y} \rangle | } },$$
where $\vec{q}$ is the flow vector of charged tracks at mid-rapidity
$$ q_x = \frac{ \sum_{j=1}^{M} w_j \cos(\varphi_j) }{ \sum_{j=1}^{M} w_j }, \quad q_y = \frac{ \sum_{j=1}^{M} w_j \sin(\varphi_j) }{ \sum_{j=1}^{M} w_j }. $$
Weights $w_j (\varphi, \eta, p_\text{T})$ are used to correct for non-uniform acceptance and reconstruction efficiency.
$ v_1$ can be further decomposed into a rapidity-odd component. Conventionally \cite{ALICEv1}, it is defined such that the directed flow of the spectator neutrons at positive pseudorapidity has positive sign
$$ v_1^\text{odd} = \frac{1}{2} ( v_1\lbrace \text{ZDC-A} \rbrace - v_1\lbrace \text{ZDC-C} \rbrace ). $$

\section{Results}
\begin{figure}[H]
  \centering
\includegraphics[width=0.5\textwidth]{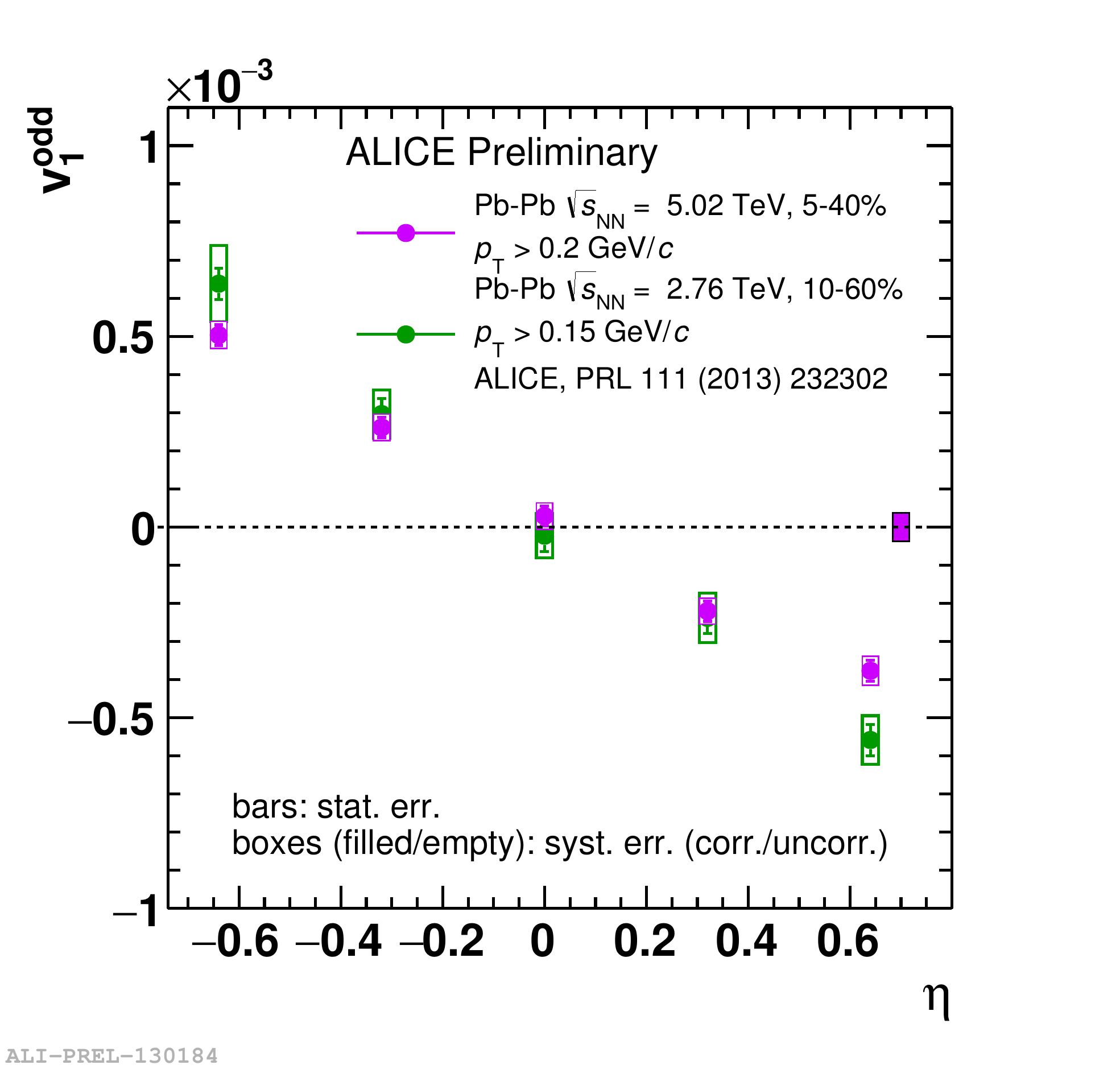}
 \caption{$v_1^\text{odd}$ of unidentified charged particles in Pb--Pb collisions at $\sqrt{s_\text{NN}} = 5.02$ and 2.76 TeV \cite{ALICEv1}.}
  \label{fig:1}
\end{figure}

Figure~\ref{fig:1} shows $v_1^\text{odd}$ of inclusive charged particles as a function of pseudorapidity and compare it with results at $\sqrt{s_\text{NN}} = 2.76$ TeV \cite{ALICEv1}. A decrease of a factor $1.3$ is observed in the slope d$v_1^\text{odd} /$d$\eta$ going from 2.76 to 5.02 TeV collision energy, which is qualitatively consistent with the energy dependence previously observed from lowest RHIC energies to LHC ones \cite{Adamczyk:2014ipa}. This can be interpreted as a decrease in the rotation of the system in the reaction plane at these higher energies. The difference in the centrality ranges used in Fig.~\ref{fig:1} for results at $\sqrt{s_\text{NN}} = 2.76$ and 5.02 TeV does not affect the comparison because no significant centrality dependence of $v_1^\text{odd}$ within $|\eta|<0.8$ has been observed at either energies.

\begin{figure}[H]
\centering
\begin{subfigure}[b]{0.49\textwidth}
\includegraphics[width=\textwidth]{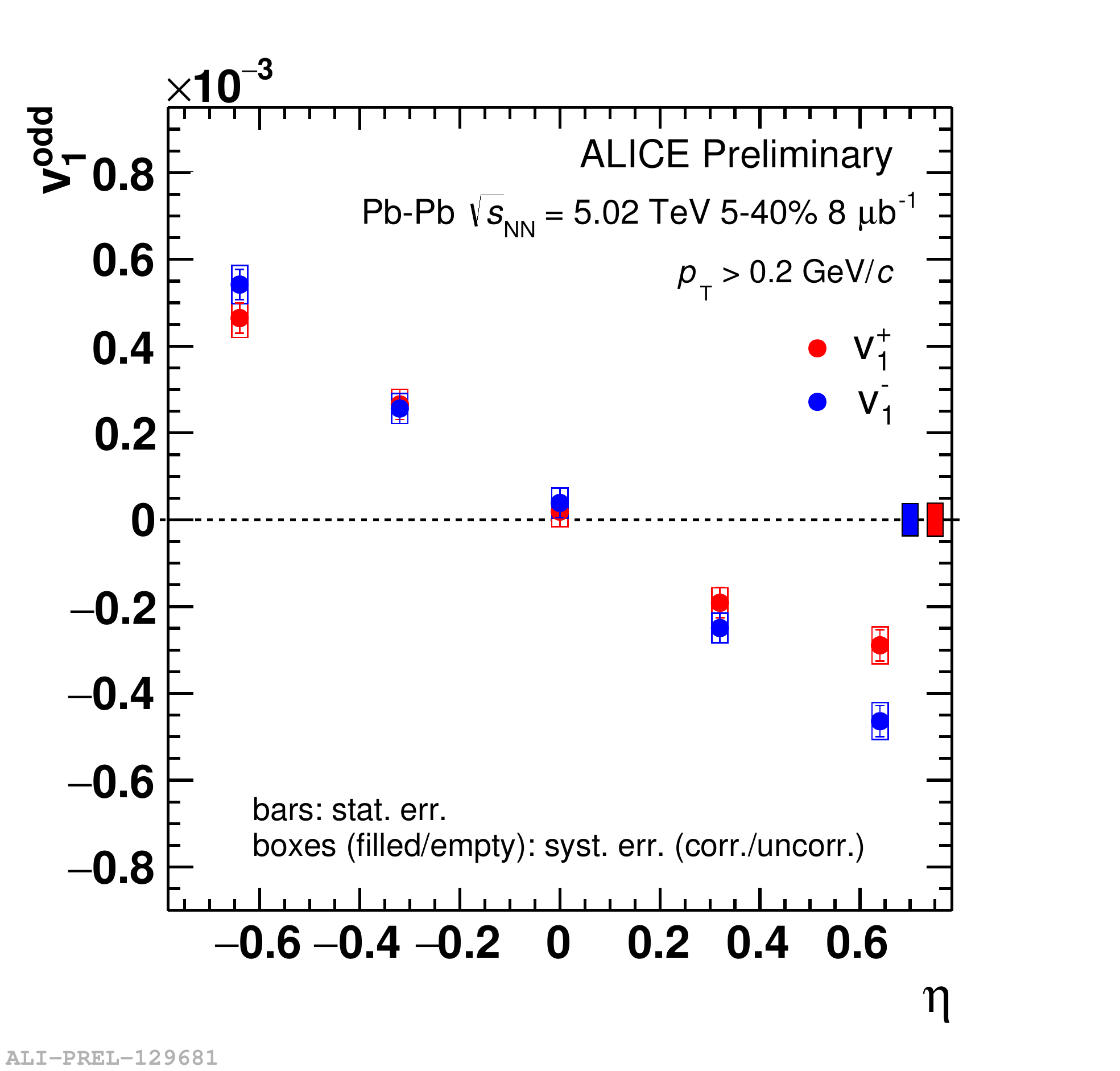}
\end{subfigure}
\begin{subfigure}[b]{0.49\textwidth}
\includegraphics[width=\textwidth]{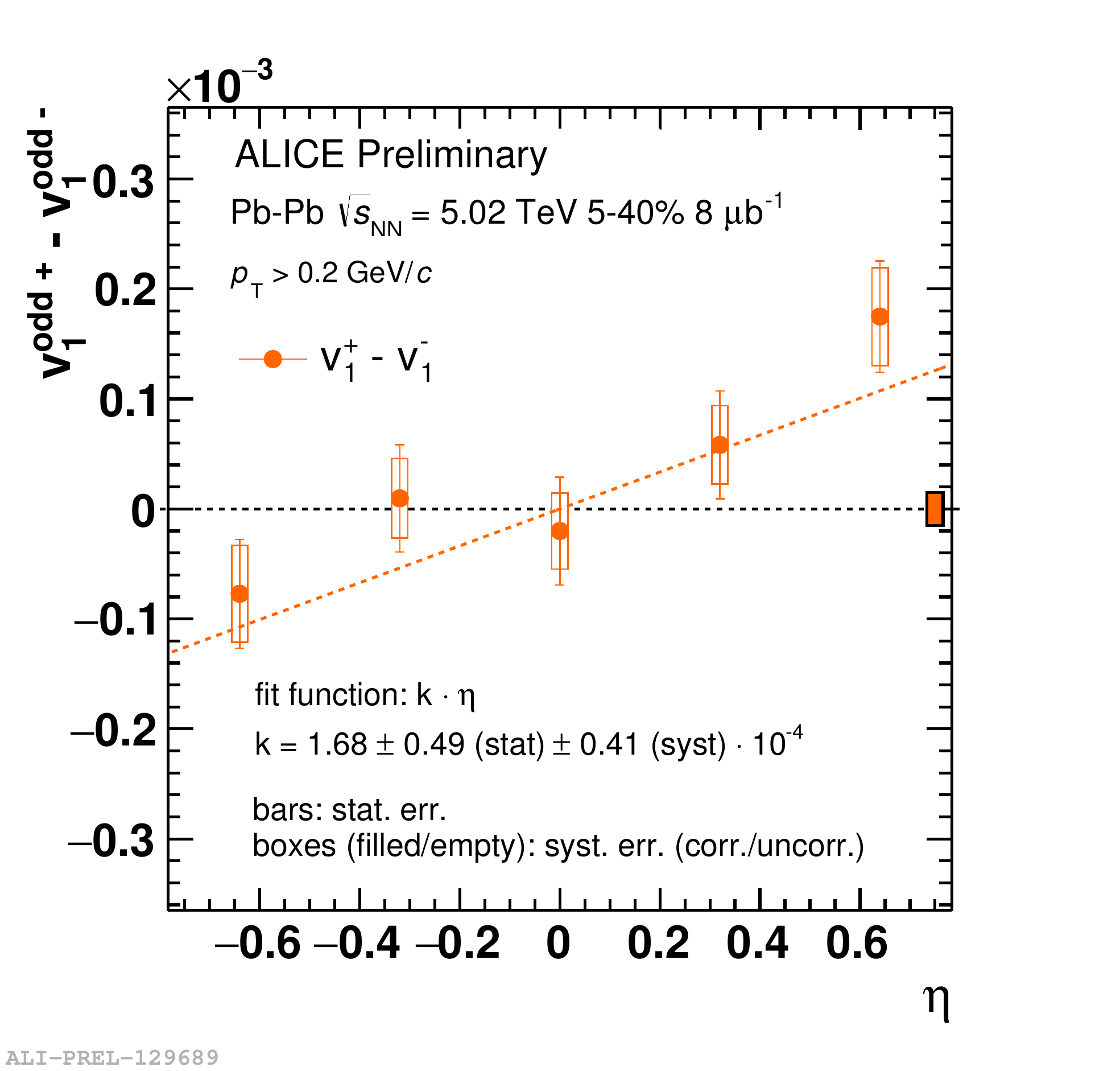}
\end{subfigure}
\caption{Left panel: $v_1^\text{odd}$ of unidentified positevely (red) and negatively (blue) charged particles in Pb--Pb collisions at $\sqrt{s_\text{NN}} = 5.02$ TeV. Right panel: charge difference $\Delta v_1^\text{odd} = v_1^\text{odd+} - v_1^\text{odd-}$, fitted with a function of the form $\Delta v_1^\text{odd} (\eta) = k \cdot \eta$.}
\label{fig:2}
\end{figure}

Figure~\ref{fig:2} shows $v_1^\text{odd}$ at $\sqrt{s_\text{NN}} = 5.02$ TeV separately for positively and negatively charged particles (left panel), together with the charge difference $\Delta v_1^\text{odd} = v_1^\text{odd+}-v_1^\text{odd-}$ (right panel). In order to quantify a possible signal, the rapidity dependence of the charge difference $\Delta v_1$ is fitted using a linear function with slope $k$
$$\Delta v_1^\text{odd} (\eta) = k \times \eta, \qquad k = 1.68 \, \pm \, 0.49 \,  (\text{stat.}) \, \pm \, 0.41 \, (\text{syst.}) \times 10^{-4}.$$
The slope $k$ has a total significance of 2.6 $\sigma$. Compared to predictions \cite{Gursoy:2014aka} for $\pi^\pm$ at $\sqrt{s_\text{NN}} = 2.76$ TeV and similar $\langle p_\text{T} \rangle$, it is up to 2 orders of magnitude bigger and of opposite sign. Such discrepancies, which are not expected to be easily explainable by the differences in particle species, collision energy or $\langle p_\text{T} \rangle$, do not allow drawing firm conclusions on the origin of the observed phenomenon.

\section{Conclusions}
In this contribution the first measurement of the pseudorapidity-odd component of directed flow $v_1^\text{odd}$ of inclusive charged particles in Pb--Pb collisions at $\sqrt{s_\text{NN}} = 5.02$ TeV is presented. Compared to results at 2.76 TeV, the slope d$v_1^\text{odd} /$d$\eta$ is observed to decrease by a factor $1.3$. A hint of a charge-dependent difference is observed, which was quantified as a difference in the slope d$v_1^\text{odd} /$d$\eta$ and was found to be $1.68 \, \pm \, 0.49 \,  (\text{stat.}) \, \pm \, 0.41 \, (\text{syst.}) \times 10^{-4}$, with a total significance of 2.6 $\sigma$. This difference, which needs to be confirmed with more data and reduced systematic uncertainties, can be attributed to the influence of the early-time magnetic field, but both the sign and the magnitude differ from predictions \cite{Gursoy:2014aka}. Therefore, more work is required on both the experimental and theoretical side to assess this interpretation; if this latter will be confirmed, these measurements could provide strong constraints on the strength and lifetime of the magnetic field generated in high-energy heavy-ion collisions.